\documentclass[12pt,preprint,tighten,flushrt]{aastex}

\usepackage{epsfig}
\usepackage{colordvi}
\usepackage{html}

\newcommand{\rtp}[1]{\ensuremath{^{#1}}}

\newcommand{\apx}{\ensuremath{\sim}}

\newcommand{\HIPPARCOS}{{\em HIPPARCOS}}
\newcommand{\LEAVITT  }{{\em LEAVITT}}
\newcommand{\HST}      {{\em HST}}

\newcommand{\DARWIN}   {{\em DARWIN}}
\newcommand{\KEPLER}   {{\em KEPLER}}
\newcommand{\COROT}    {{\em COROT}}
\newcommand{\TPF}      {{\em TPF}}

\newcommand{\GAIA}     {{\em GAIA}}
\newcommand{\FAME}     {{\em FAME}}
\newcommand{\AMEX}     {{\em AMEX}}
\newcommand{\OBSS}     {{\em OBSS}}

\newcommand{\etal}{{\it et al.}}

\newcommand{\ad} {\ensuremath{^{\rm o} \;}}

\newcommand{\snROv}[1]{\renewcommand{\baselinestretch}{#1}\begin{normalsize}}
\newcommand{\enROv}{\end{normalsize}\renewcommand{\baselinestretch}{1.0}}

\newcommand{\ssROv}[1]{\renewcommand{\baselinestretch}{#1}\begin{small}}
\newcommand{\esROv}{\end{small}\renewcommand{\baselinestretch}{1.0}}

\begin{document}

\title{{\Large LEAVITT}\\ 
\vspace*{1cm}
A MIDEX-class Mission for Finding \& Characterizing \\
10,000 Transiting Planets in the Solar Neighborhood\\
\vspace*{1em} {\small A White Paper for the Exo Planet Task Force}}

\author{\vspace*{1cm}
        {\Large March 2007}\\
        \vspace*{1cm}
        Rob P. Olling\rtp{1}}

\affil{
\rtp{1}Department of Astronomy, University of Maryland at College Park\\
       email: olling@astro.umd.edu
}
\section*{}

\maketitle

\clearpage

\setcounter{page}{1}

 \section*{Abstract}
\label{sec:Abstract}
\vspace*{-0.9em}

We propose a MIDEX-class space mission with the goal to find and
characterize roughly 10,000 transiting planets. When transits occur, a
much more detailed characterization of the planet is possible
(\S\ref{sec:Introduction}), and so a large data base of transiting
planets will provide planets with a large range in periods and radii
for follow-up studies. Our survey will be all-sky and focused on stars
brighter than V=14.8. Down to V=12, \LEAVITT\ will be able to detect
Neptune-sized objects. Because of it's high cadence, \LEAVITT\ is
about 100 times more sensitive at detecting transits than \GAIA, while
it will find more than 20 times as many transits as \KEPLER. \LEAVITT\
has multi-band photometric capability implemented via a low-res
dispersive element which can obtain 0.2\% (2 mmag) photometry down to
V=14.8. \LEAVITT's high multi-band photometric accuracy reduces the
number of false-positives significantly.

\vspace*{-2.5em}
  \section{Introduction}
\label{sec:Introduction}
\vspace*{-0.9em}
Just over two hundred extra-solar giant planets (ESGPs) are currently
known in 176 planetary systems\footnote{
http://vo.obspm.fr/exoplanetes/encyclo/encycl.html and
http://exoplanets.org/
} \citep{Bea2006}. Statistical analysis of these detections indicates
that roughly 10\% of ``Sun-like'' (main-sequence of type F, G \& K)
stars have fairly massive planets. \citet{TT2002} determined the
probability density function ($PDF$) for ESGPs as a function of mass
and period for the then 69 known ESGPs.. They find that 3.5\% of stars
have planets in the period ($P$) and mass ($M$) range of: $2\,
\rm{days} \le P \le 10\, \rm{years}$ and $1 \le M \le 10$ Jupiter
masses ($M_J$).  We use a scaled-up version (by a factor of 1.62) to
account for the current, higher normalization \citep{S2005}. Note that
the TT2002 $PDF$ ($PDF_{ESGP}$) diverges for large periods {\em and}
small masses, so that it is likely that the $PDF_{ESGP}$ turns over at
both low masses and long periods. The extrapolation of the TT2002 PDF
to Uranus/Neptune masses increases the numbers by a factor of 2.6, and
we find that 16\% of stars should have planets in the range $2\,
\rm{days} \le P \le 10\, \rm{years}$ and $0.05 \le M \le
13\, M_J$. Extrapolating to Earth-mass planets and up to the period of
Neptune, the state-of-the art $PDF_{ESGP}$ predicts planets around two
out of three stars. \\
\hspace*{1.4em}
In two related proposals, we propose to explore two extremes of the
PDF: 1) long-period massive planets (``Finding Solar System Analogs
With SIM and HIPPARCOS'') and 2) Earth-mass planets in the habitable
zone (``Hunting for Earth-Mass Extra-Solar Planets with the Dispersed
Fourier Transform Spectrometer''). Here we propose a MIDEX-class,
space-based survey aimed at finding and characterizing about 10,000
transiting planets. Finding transiting planets is not so difficult as
they periodically dim the light of their parent stars substantially
(at the 1\% level). However, {\em knowing} that an observed dimming is
due to the transit of a planet rather than to myriad other possible
effects (false-positives), that is the hard part. In fact, it is
though that the false-positives outnumber planetary transits (PTs) by
about a factor of one hundred. We will circumvent this problem to a
large degree by looking at bright sources to avoid confusion, and by
obtaining simultaneous multi-color photometry at the milli-magnitude
level. See \S\ref{sec:MIDEX-Class_LEAVITT_Mission} below for a
detailed description of our mission concept and implementation.\\
\hspace*{1.4em}
In contrast to radial-velocity (RV) of astrometric surveys, a transit
survey will not yield the masses of the planets, but the otherwise
inaccessible planetary radius. Since the masses can be determined via
RV follow up (and the known inclination from the transits), our
proposed survey would provide the mean density for a large number of
extra-solar planets. Also, these planets allow for the search for and
detection of planetary atmospheres via transmission spectroscopy
\citep{CBNG2002} while the combination of on- and off transit
spectroscopy can also yield the continuum [e.g., \citet{DHSR2006,
Cea2005}] and emission spectra \citep{RDHSH2007, Spitzer2007}.  Both
detections and non-detections of continuum and spectral features in
the spectra of ESGPs lead to improved knowledge of their atmospheric
properties such as temperature, albedo, dust contents, cloud cover,
heat redistribution, weather and so forth [e.g., \citet{RDHSH2007,
SRHMCD2005, Fea2003, MCSH2003, BLC2002}].  Also, eclipse-timing
techniques \citep{I1959} can be used to search for additional (down-to
Earth-mass) planets [e.g., \citet{AS2007, ASSC2005, SA2005, ME2002}]
in the most suitable systems.\\
\hspace*{1.4em}
Thus, to quote \citet{CBBL2007}, ``When extrasolar planets ... transit
their parent stars, we are granted unprecedented access to their
physical properties.'' In fact, spectroscopy of extra-solar planets
could eventually lead to the detection of life on another planet
[e.g., \citet{Tea2006, T2006, SF2005,DM2002}], which is of course the
goal of NASA's \TPF\ missions\footnote{
http://planetquest.jpl.nasa.gov/TPF/tpf\_index.cfm
} and ESA's \DARWIN\footnote{
http://sci.esa.int/science-e/www/area/index.cfm?fareaid=28
} project.\\
\hspace*{1.4em}
Finally, it has been long presumed that the migration of giant planets
towards close-in orbits would destroy the proto-planetary disk and
would thus inhibit the formation of additional planets. However, the
most recent simulations indicate that while the migration temporarily
destroys the accretion process, the formation of earth-mass planets
goes on in about 60\% of the models studied by \citet{FN2007, FN2005}
and \citet{MRS2007}.  This perhaps surprising result can be explained
by the fact that giant-planet formation+migration occurs very rapidly,
as compared to the formation of terrestrial planets. As a result, the
``perturbation'' due to migration is minor.  These results indicate
that our proposed survey of transiting planets would also yield a
catalog of systems that would be enriched with earth-mass planets
(with respect to blind searches).

\vspace*{-2.5em}
  \section{Transit Surveys}
\label{sec:Transit_Surveys}
\vspace*{-0.9em}
There is a large number of on-going ground-based transit surveys
inspired by the apparent ease of detecting the \apx1\% transit signal
[e.g., \citet{H2003}]. However, the results have been at least an
order of magnitude smaller than expected. To quote \citet{PZQ2006},
``... Altogether, thousands of [1--4 m] telescope nights have been
invested in these surveys, monitoring hundreds of thousands of target
stars in the solar neighborhood and in the Galactic disc. However,
even after years of operation, the results of these surveys failed to
meet the expectations, with only a slow trickle of detections instead
of the expected bounty. ...'' end quote. \citet{PZQ2006} argue that
this is due to correlate noise on time-scales of several hours that
are due to, for example, atmospheric effects, temperature, and
tracking errors vary on roughly the same time-scale of several hours:
the duration of the transit \citep{P_ISSI2006}. In fact, it is (or
should have been) rather well known that milli-magnitude photometry
is rather difficult to achieve from the ground. This is exactly the
reason why \HIPPARCOS\ \citep{ESA97} photometry is the best available,
and why \GAIA\footnote{
http://gaia.esa.int/science-e/www/area/index.cfm?fareaid=26
} spends so much of its focal plane of photometry (and
spectroscopy). For example, the extremely well-calibrated {\em SDSS}
survey achieves roughly 1\% (\apx10 mmag) relative photometry
\citet{P_SDSS2007}.

\vspace*{-2.5em}
  \section{Mission Concept}
\label{sec:Mission_Concept}
\vspace*{-0.9em}
On the other hand, as stated above, photometry at the mmag level is
much more easily achieved with space-based platforms. However, note
that for example \HST\ photometry is not that accurate. To achieve
this goal, one must repeatedly observe the same star many times, while
it also crucial to have excellent knowledge of the point-spread
function (PSF). The former criterion is required so as to be able to
remove long-term trend, eliminate instrument-related systematics
etc.. In other words, demand consistency. Knowledge if the PSF is
crucial because (in crowded fields) PSF-fitting yields superior
integrated magnitudes. If a mismatch exists between actual PSF and
assumed PSF, systematic effects will creep in the photometry. For this
reason, astrometric programs are well-suited to obtain very accurate
photometry, because they too need many repeat measurements and
exquisite PSF control/knowledge such as in \HIPPARCOS, \GAIA\
\citep{GAIA} and the canceled \FAME\ project \citep{FAME} and the
proposed \AMEX\ \citep{AMEX_2005, AMEX_2003} and \OBSS\ missions
\citep{OBSS}.

\vspace*{-2.5em}
\subsection{The MIDEX-Class LEAVITT Mission}
 \label{sec:MIDEX-Class_LEAVITT_Mission}
\vspace*{-0.9em}
Based on our extensive experience with proposed the \FAME, \AMEX\ and
\OBSS\ astrometric missions, and their transit capabilities in
particular [e.g., \citep{RPO2003, OG2002, GO2002}], we propose the
following \LEAVITT\footnote{
We propose this project in honor of Henrietta Swan Leavitt who
contributed very significantly to precision astrophysics with the
discovery of Cepheid variables in the Magellanic Clouds almost one
hundred years ago, and opened up the field of temporal
astrophysics. Alternatively, LEAVITT could stand for: ``LEgacy
Astrophysics of Variable, Intermittent and Transiting Things.''
} MIDEX-class space mission.\\
\hspace*{1.4em}
The basic property a transit survey needs to have is a rapid cadence,
therefore, \LEAVITT\ will spin every 90 minutes. A ``precession
period'' of 15 days ensures that 70\% of the sky is observed about
once every week (the remaining 30\% is inaccessible due to the Sun
exclusion zone).  In total, the number of observations will be 10,500
in 5 years. \\
\hspace*{1.4em}
So as to keep the mission with the 159 M\$ MIDEX budget, we propose a
(cheap) Sun-synchronous Polar orbit such as those of IRAS,
COROT\footnote{
http://smsc.cnes.fr/COROT/
} and many others. Like the scanning astrometric missions, \LEAVITT\
would have two viewports separated by a basic angle of \apx90\ad that
project the light onto the same focal plane. However, significant cost
reductions are achieved because we are dealing with a photometric
mission, so that we do not require the exquisite basic angle stability
as required by \LEAVITT's astrometric cousins. The mirror is
rectangular and measures 14x55 cm with a focal length of 5 meters. \\
\hspace*{1.4em}
The instantaneous field of view is 3.5\ad x 3.5\ad, which is covered
by 36 CCD (5,120 x 5,120 with 10 $\mu$ pixels (similar to \OBSS) that
operate in drift-scan mode (like SDSS and \GAIA, etc.). The wide field
of view ensures that a given stars is observed regularly for about 6.6
hours before the scanned strip moves off target. Typically, this
period (epoch) is long enough to cover a planet transiting its parent
(2 to 3 hours), as well as have baseline observations outside
transit.\\
\hspace*{1.4em}
A crucial aspect of the instrument is the inclusion of a dispersive
element that creates rather low-res ($R\sim100$) slitless ``spectra''
for each object. These spectra are required for two reasons: 1) it
extends the dynamic range of the instrument by roughly 5 magnitudes,
and 2) it provides highly accurate color information that is crucial
for the characterization of the transit event, and hence reduce the
false-alarm rate to manageable levels. \citet{T2004} shows that color
changes during transit are at the 1 mmag level, while false-positive
are several times larger.  Other, color-independent methods can also
be used to reduce the false-alarm rate [e.g., \citep{TS2005,
SMO2003}]. Our usable magnitude range is 5.7 (saturation) to 14.8 (2
mmag photometry for 3-color photometry per hour integration time).\\
\hspace*{1.4em}
The moderate dispersion can be achieved by a low-power dispersive
element prism in the light-path (such as was planned by {\em
DIVA}\footnote{http://www.ari.uni-heidelberg.de/diva/} or \COROT, or
ESO's WIFI imager)
\\
\hspace*{1.4em}
The above strategy implies that \LEAVITT\ will visit the average star
\apx168 times in 5 years, while \apx61 observations are taken during
those 168 6.6-hour epochs.\\
\hspace*{1.4em}
We simulated the expected number of transits in the following manner:
1) we perform a full-length mission simulation that measures the
number of visits at arbitrary points on the sky, 2) these data are
then used to determine a sky-averaged probability for detecting {\em
five} transit-like events with a duration determined by the period of
the planet (transits last longer in longer-period orbits), 3) here we
assume a G2V primary and a planet with Jupiter's radius, 4) we use the
modified TT2002 $PDF$, multiplied this by the probability that the
planet is seen edge-on, and the probability of observing 5 transits as
determined in (2) above, 5) we use a simple star-count model of the
solar neighborhood to predict how many dwarf stars are with our
magnitude limit, and 6) generate the total number of observable
planetary transits. \\
\hspace*{1.4em}
The combined effects of \LEAVITT\ not observing continuously, and the
period distribution of the ESGPs indicates that \LEAVITT\ is $\ga
50$\% complete for periods shorter than 18 days. This compares very
well with 2.4 days for \GAIA\ (for which we performed the same
simulations). In fact, \LEAVITT\ outperforms \GAIA\ by a factor of 100
(460) at a period of 10 (20) days. This is mostly due to \GAIA's very
slow precession rate of 75 days. \LEAVITT\ will produce roughly 20
times more PTs than \KEPLER. On the other hand, \KEPLER\ can find
planets with a radius of the Earth at $V=12$, while \LEAVITT\ can
detect Neptune-size objects at that magnitude.
\\
\hspace*{1.4em}
If we assume planets down to Uranus/Neptune mass, we expect to find
roughly 21,000 planetary transits with periods roughly up-to 30
days. The brightest subset of \apx10,000 planets will have 3-color
photometry at the 2 mmag level, and these are the systems that can
most likely be classified photometrically as transiting planets.

\vspace*{-2.5em}
\subsection{Cost Estimate}
 \label{sec:Cost_Estimate}
\vspace*{-0.9em}
We compared our \LEAVITT\ mission with the \FAME, \AMEX and \OBSS\
missions to arrive at a reasonable cost estimate. Our cost model takes
into account the overall weight, launch costs, mirror size, number of
CCDs, electronics, instrument weight, bus mass, attitude control and
science operations. We define scaling relations that result in good
estimates for the \FAME, \AMEX and \OBSS\ budgets. This model results
in a total cost for the \LEAVITT\ mission of 159 M\$ (FY2005).

\snROv{0.90}

\begin{flushleft}

\end{flushleft}

\enROv

\end{document}